\documentclass{article}
\usepackage{spconf,amsmath,graphicx}
\usepackage{cite}
\usepackage{float}
\usepackage{graphicx}
\usepackage{epsf}
\usepackage{array}
\usepackage{makecell}
\usepackage{amsmath}
\usepackage{amssymb}
\newcolumntype{M}[1]{>{\centering\arraybackslash}m{#1}}
\usepackage{lineno}
\usepackage{fixmath}
\usepackage{hyperref}
\usepackage{hyperref}

\title{EVALUATION OF DEEP-LEARNING-BASED VOICE ACTIVITY DETECTORS AND ROOM IMPULSE RESPONSE
MODELS IN REVERBERANT ENVIRONMENTS}
\name{Amir Ivry \qquad Israel Cohen \qquad Baruch Berdugo \thanks{This work was supported by the Israel Science Foundation (grant no. 576/16) and the ISF-NSFC joint research program (grant No. 2514/17).}}
\address{Andrew and Erna Viterbi Faculty of Electrical Engineering \\
Technion -- Israel Institute of Technology, Technion City, Haifa 3200003, Israel}

\begin{document}
%
\maketitle
\begin{abstract}
State-of-the-art deep-learning-based voice activity detectors (VADs) are often trained with anechoic data. However, real acoustic environments are generally reverberant, which causes the performance to significantly deteriorate.
To mitigate this mismatch between training data and real data, we simulate an augmented training set that contains nearly five million utterances. This extension comprises of anechoic utterances and their reverberant modifications, generated by convolutions of the anechoic utterances with a variety of room impulse responses (RIRs).
We consider five different models to generate RIRs, and five different VADs that are trained with the augmented training set.  We test all trained systems in three different real reverberant environments.
Experimental results show $20\%$ increase on average in accuracy, precision and recall for all detectors and response models, compared to anechoic training. Furthermore, one of the RIR models consistently yields better performance than the other models, for all the tested VADs. Additionally, one of the VADs consistently outperformed the other VADs in all experiments.
\end{abstract}
\section{Introduction}\label{sec:intro}
Voice activity detection (VAD) aims to determine the boundaries in which speech exists in an observed audio signal. State-of-the-art deep-learning-based VADs are often trained with anechoic data. However, real-life acoustic environments are reverberant, which deteriorates VAD performance in practical scenarios. In this study, we mitigate the mismatch between training data and real data by generating an augmented training set that integrates anechoic and reverberant audio signals. The reverberant training corpus is generated by convolving anechoic utterances with simulated room impulse responses (RIRs). Enhanced VAD in reverberant environments may benefit a variety of audio-based applications such as speech enhancement \cite{kinoshita2016summary, kuklasinski2016maximum, zhao2016dnn}, dereverberation \cite{han2015learning, schwarz2015coherent} and speech and speaker recognition \cite{ko2017study, giri2015improving}.

Deep-learning-based VADs have attained leading performances during recent years, due to the ability of neural networks to learn non-linear relations and complex patterns of audio signals. To detect voice activity, Ariav and Cohen \cite{ariav2018deep} encoded spectral audio features via an auto-encoder that fed a recurrent neural network. Wagner et al. \cite{wagner2018deep} introduced automatic feature engineering through the convolutional layers of a deep neural network. Leading performance was obtained by Kim and Hahn \cite{kim2018voice} that integrated an attention model to weight context information into existing deep learning architectures. Combined end-to-end VAD system was introduced by Ariav et al. \cite{ariav2019end}, that comprised of WaveNet for feature extraction and a deep residual network for speech detection. Ivry et al. \cite{ivry2019voice} applied ensemble learning with two deep encoder-decoder structures to learn the unique temporal and spatial patterns of speech through the diffusion maps method.

In latest decades, several RIR models were proposed to produce reverberant utterances via simulations. An extension of the known image method \cite{allen1979image} to arbitrary polyhedra was first introduced by Borish \cite{borish1984extension}. Vorl\"{a}nder \cite{vorlander1989simulation} suggested a combined modeling that considers both the image method and ray-tracing techniques. Rindel \cite{rindel1993modelling} employed reflection coefficients that are incidence angle-dependent in the frequency domain, to offer a more accurate characterization of a room response. A similar model was implemented by Lam \cite{lam2005issues}, but it focused on low frequencies for more realistic boundary conditions. Valeau et al. \cite{valeau2006use} applied the diffusion equation to predict room acoustics.

We consider the aforementioned five deep-learning-based VADs \cite{ariav2018deep,wagner2018deep,kim2018voice,ariav2019end,ivry2019voice} and five RIR models
\cite{borish1984extension,vorlander1989simulation,rindel1993modelling,lam2005issues,valeau2006use}. First, we show that training these detectors with solely anechoic corpus and testing them in real reverberant rooms and spaces leads to a significantly impeded detection capability. To include unique patterns and acoustic features of reverberant data during training, we generated an augmented training set of nearly five million utterances. This extended corpus comprises of anechoic and reverberant signals, where the latter is generated by convolving the anechoic signals with a variety of RIRs, generated using a fixed RIR model. Then, all five VADs are independently trained with this augmented training set. This experiment is repeated for each of the five RIR models. All trained detection systems are tested in three real reverberant spaces of a classroom, a large concert hall, and an octagon shaped library. Experimental results demonstrate that the performance of all detectors is enhanced in each of the tested reverberant environments, regardless of the RIR model employed during training. Evaluation measures such as accuracy, precision and recall increase by $20\%$ on average, compared to non-reverberant training. An interesting outcome shows in each of the tested setups, the leading accuracy of each detector was consistently achieved by the Valeau RIR model \cite{valeau2006use}. In a similar manner, the detector introduced in Ivry \cite{ivry2019voice} prevailed competing VADs across all experiments.

The remainder of this paper is organized as follows. In Section \ref{sec:database}, we describe the database. In Section \ref{sec:results}, experimental results are given. We conclude in Section \ref{sec:conclusions}.

\section{Database Generation} \label{sec:database}
In this section, we detail the construction of two disjoint datasets: An augmented training set and a test set. The training set contains both anechoic and reverberant utterances, that are generated by simulating a fixed RIR model and convolving the anechoic data with it. In contrast, the test set is constructed with real reverberant conditions, not simulations.

For the training stage, we employ the TIMIT \cite{garofolo1988getting} training dataset that contains 4620 anechoic utterances, sampled at 16~kHz. Since this corpus is imbalanced and does not comprise of noises, we perform several preprocessing steps. Initially, since in TIMIT there are more speech frames than silence frames, we manually add 2~s of silence for each existing recording in the corpus. Next, we acquire recordings of stationary noises such as white and colored Gaussian noise, musical instruments and babble. These noises are randomly added to both speech and silence frames in signal-to-noise-ratios (SNRs) that are distributed uniformly between 10-20~dB relative to clean anechoic speech.

We perform augmentation of this anechoic training set, so it holds both anechoic and simulated reverberant data. To simulate varied reverberant environments, $50$ rectangular spaces are considered, such that the length, width and height are uniformly chosen from the range $3-20$~m. This permits both small, medium and large spaces. To cover various scenarios, each of the $50$ spaces is simulated $20$ times, with different locations of the speaker and the receiver. To obtain a realistic setting, the speaker and the microphone are limited to height range of $1-2$~m, and a distance of at least $0.5$~m from each other. Each room is simulated with a reverberation time (RT60) that is chosen uniformly from the interval $0.1-1$~s, such that both low and high reflective surfaces are accounted for.

Given an RIR model, we simulate $50\times20$ RIR signals. Each of these responses is convolved with the anechoic utterances in \cite{garofolo1988getting}, which results in a reverberant training set. The augmented training set is simply a composition of the original anechoic signals with their aforementioned reverberant modifications. Ultimately, for a given RIR model, the training set comprises of $4620\times1001$ utterances.

In the test stage, we use $100$ anechoic utterances from the TIMIT test dataset. To obtain the reverberant test set, convolution is applied between this corpus and real recordings of room responses. These RIRs are taken from three reverberant environments \cite{stewart2010database} of a classroom, a large concert hall and an octagon shaped library. For each environment, $130$ recordings are available, from various locations in the room. Thus, three test sets are formed, each comprises of $100\times130$ reverberant utterances.

\section{Experimental Results} \label{sec:results}
In the following experiments, voice activity detection performance is evaluated by several measures. The receiver operating characteristic (ROC) curve is used to present a trade-off between speech detection and false-alarm rates in various operation points. The robustness of the VAD and the sensitivity of its classifier to noises is derived by the area under curve (AUC) measure. Accuracy, precision, recall and F1-score \cite{powers2011evaluation} are also employed in this study. When combined, all measures strongly indicate on the accuracy, generalization and robustness abilities of the detector.

In this study, we consider five VADs \cite{ariav2018deep,wagner2018deep,kim2018voice,ariav2019end,ivry2019voice} and address them as Ariav-R, Wagner, Kim, Ariav-W and Ivry, respectively. Also, we employ five RIR models
\cite{borish1984extension,vorlander1989simulation,rindel1993modelling,lam2005issues,valeau2006use}, and refer them as Borish, Vorl\"{a}nder, Randel, Lam and Valeau, correspondingly. We perform the following experiment, comprises of two-stages; training stage and test stage. In the first part, a fixed RIR model is simulated. Then, the steps described in Section \ref{sec:database} are implemented with respect to the chosen RIR model. As a result, an augmented training set is obtained. Next, a VAD system is chosen and trained with the derived training set. We repeat this experiment for each VAD system and for each RIR model. Ultimately, this stage yields $5\times5$ trained VAD systems.
In the second stage, we test each trained detector on three test sets, generated in three reverberant environments of a classroom, a large concert hall and an octagon shaped library, as detailed in Section \ref{sec:database}. An experiment conducted by the authors of this study showed that these three acoustic spaces are characterized by long ($1$~s), medium ($0.8$~s) and short ($0.6$~s) reverberation time, in correspondence.

\begin{figure}[htp]
	\centering\includegraphics[width=0.88\columnwidth]{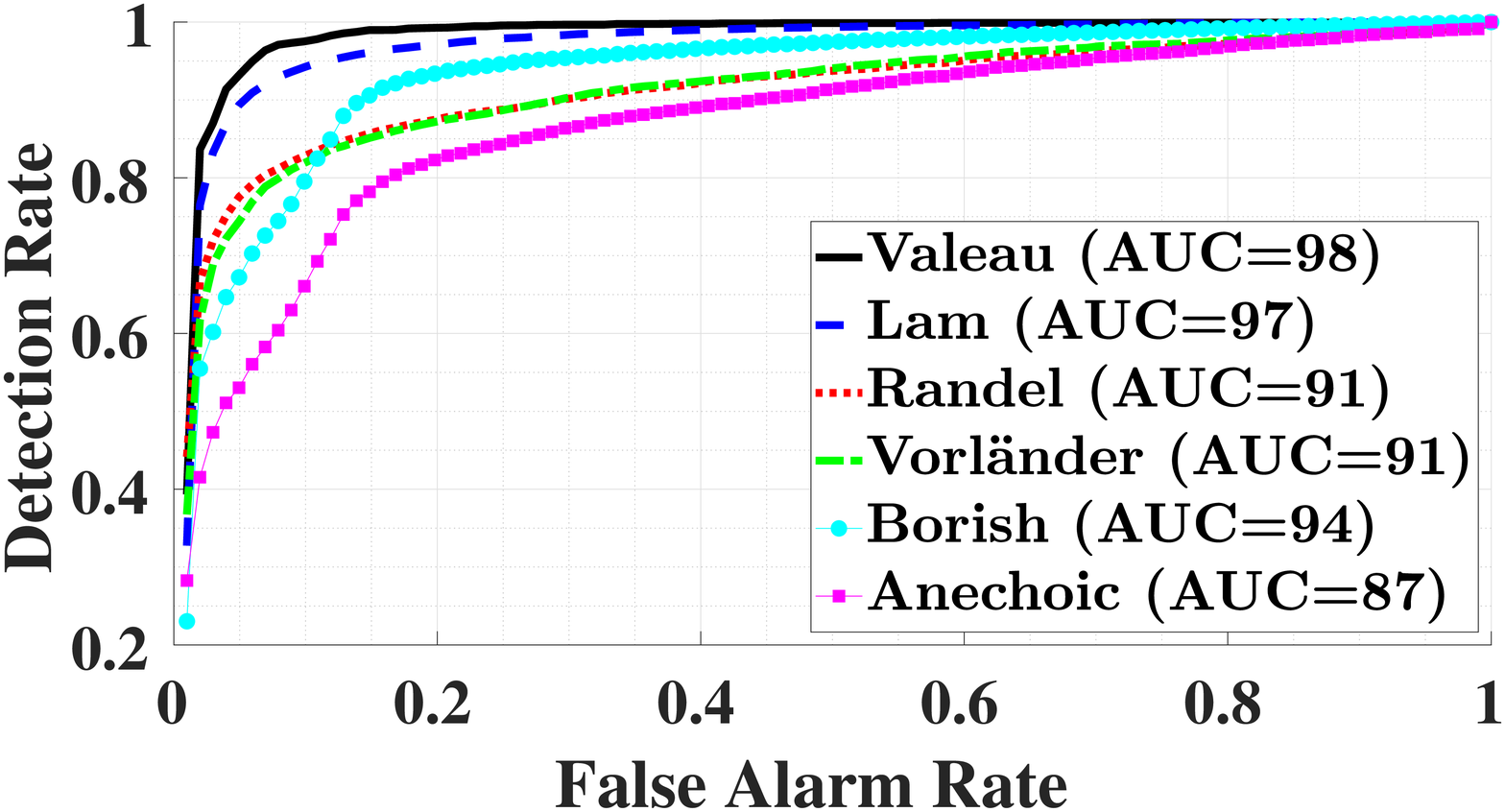}
	\vfill
	\includegraphics[width=0.88\columnwidth]{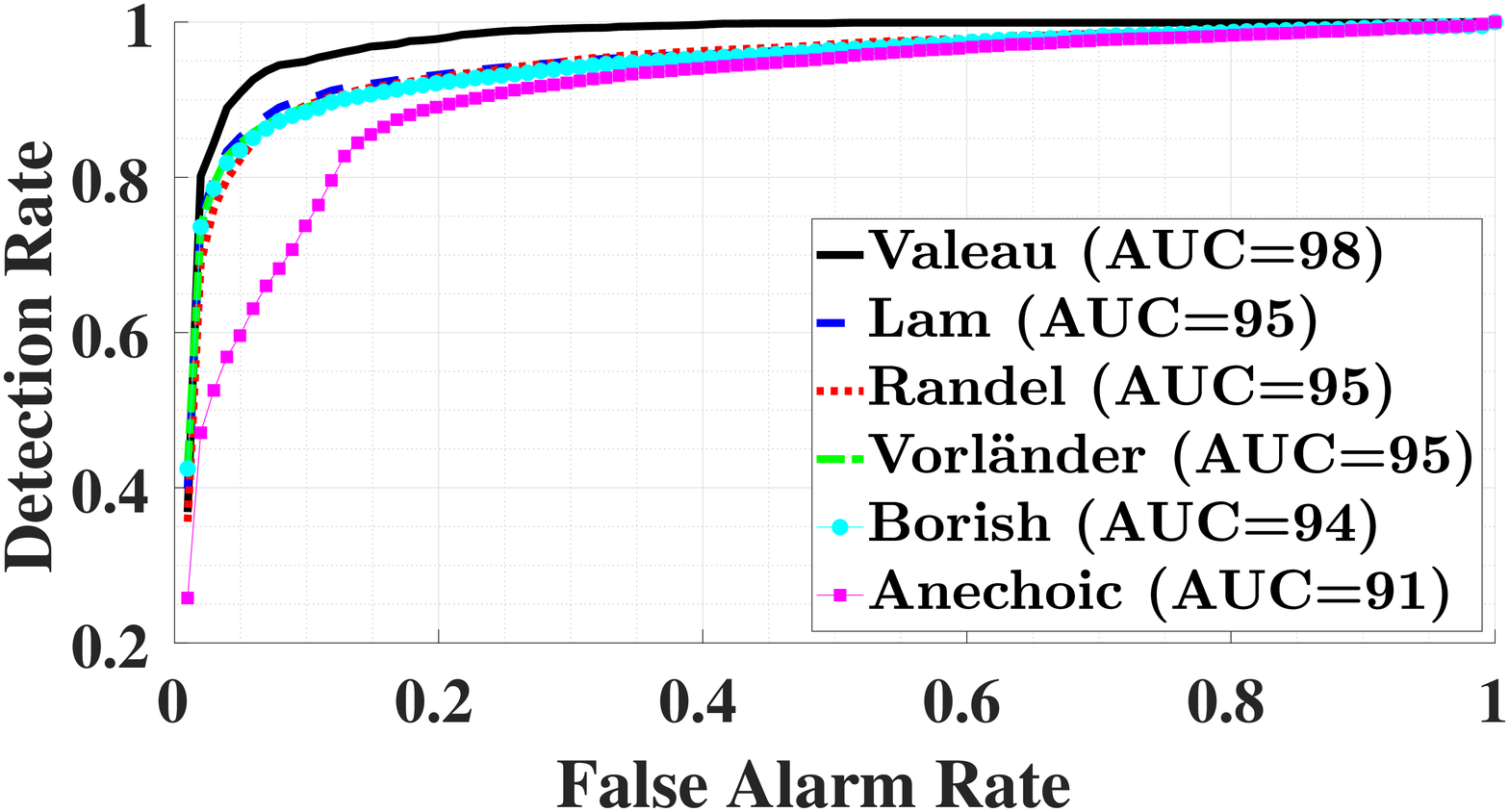}
	\vfill
	\includegraphics[width=0.88\columnwidth]{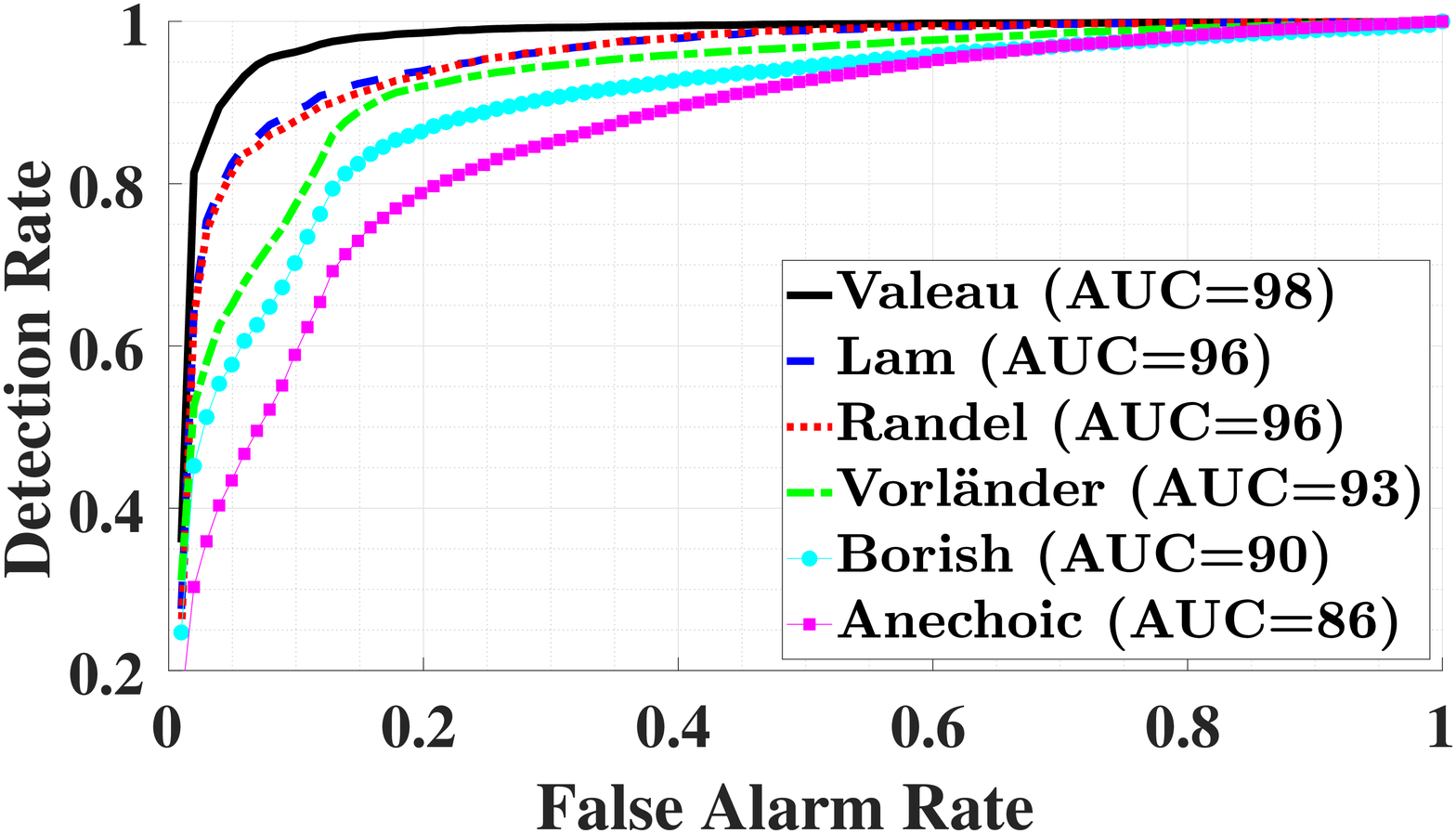}
	\vfill
	\includegraphics[width=0.88\columnwidth]{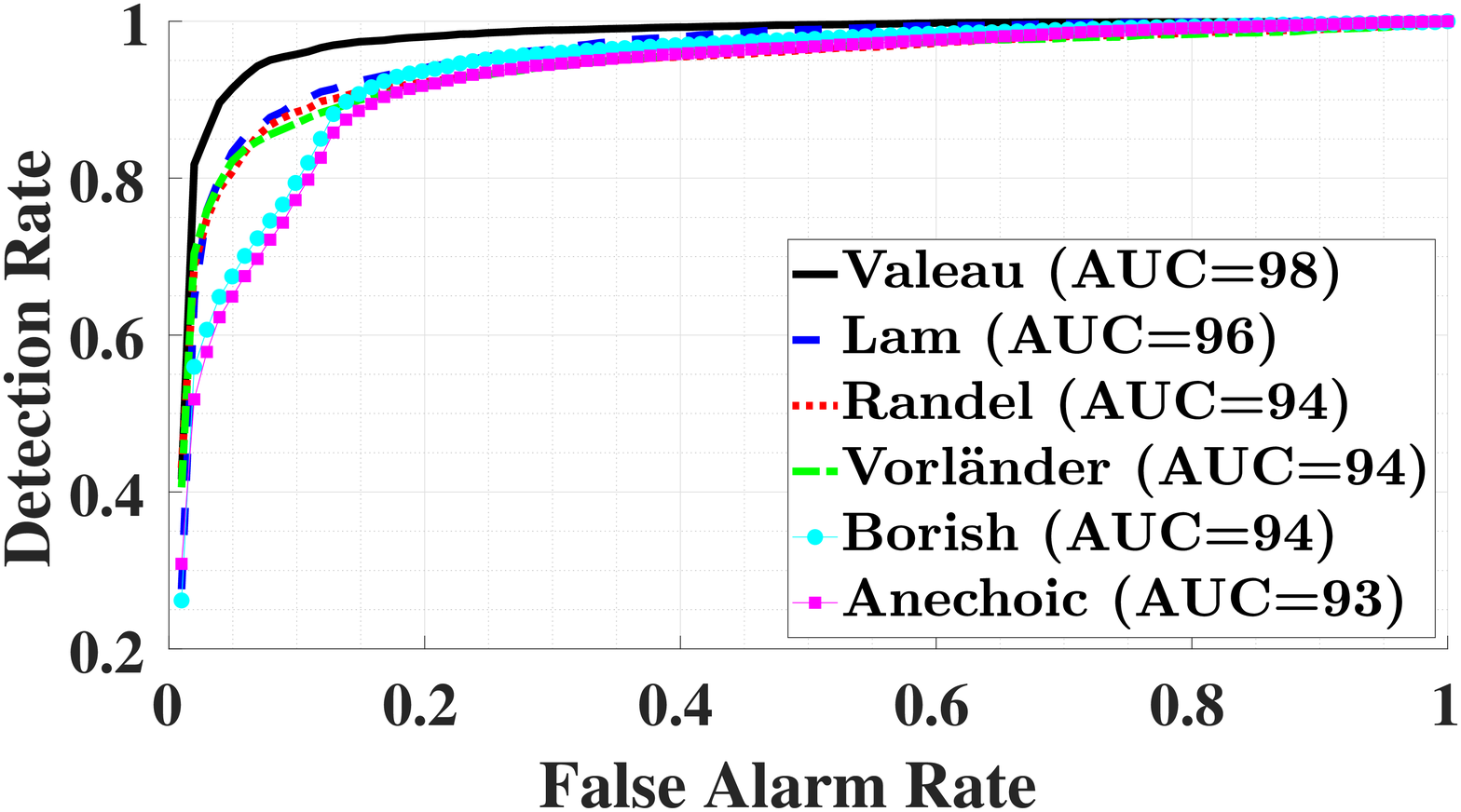}
	\vfill
	\includegraphics[width=0.88\columnwidth]{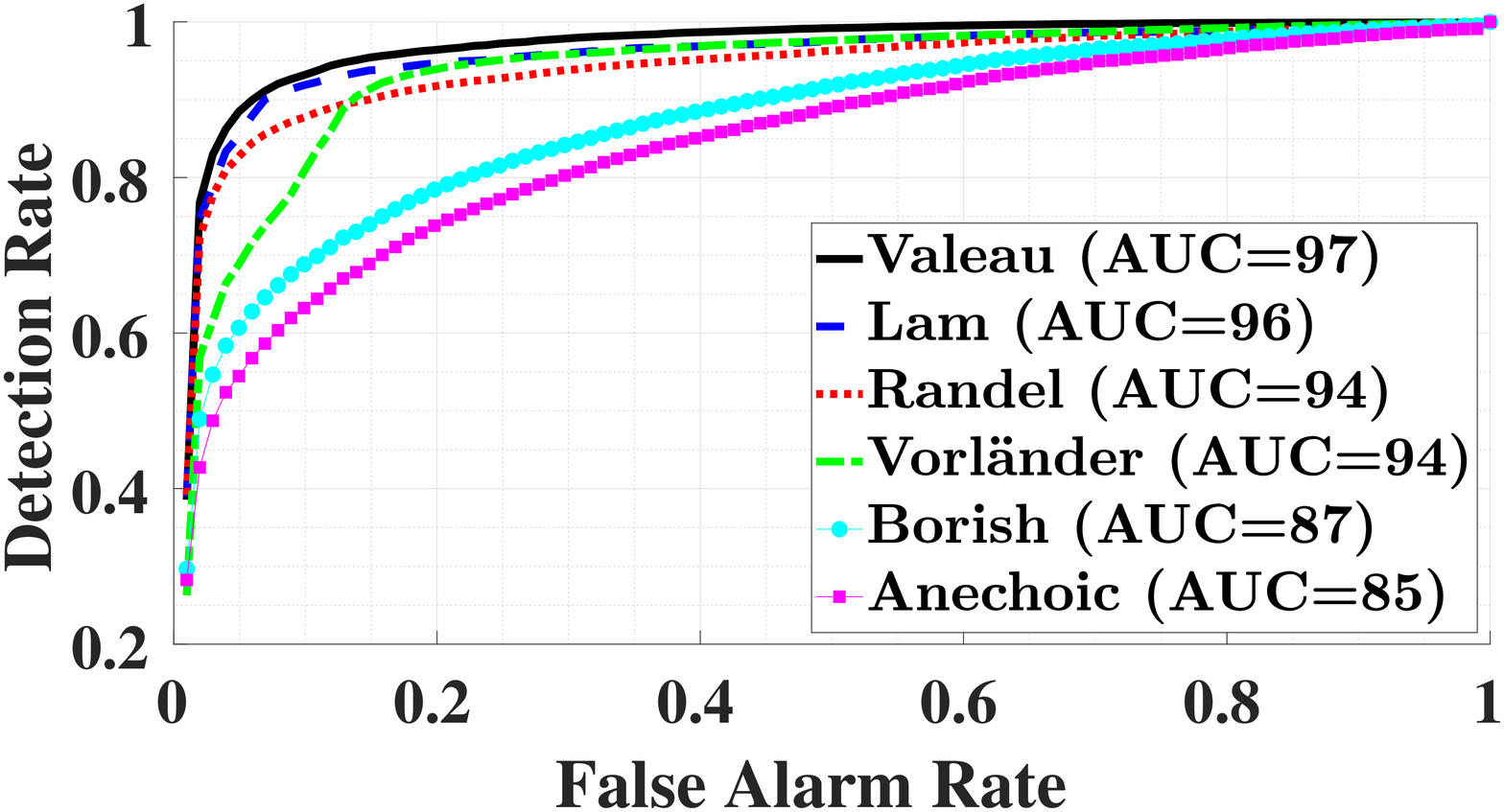}
	\caption{Detection rate versus false alarm rate in a reverberant setup of a classroom. Comparison is made between the five different training RIR models. VADs (top to bottom): Ivry, Ariav-W, Kim, Wagner and Ariav-R.}
	\label{ROCs}
\end{figure}

By observing Fig. \ref{ROCs}, several conclusions can be derived. First and foremost, if the training set contains merely anechoic data, then the performance of all VADs is significantly degraded when tested in real reverberant conditions. Respectively, employing the suggested augmented training set that comprises of reverberant utterances consistently enhances VAD performance in practical scenarios. The reason is that acoustic patterns and features highly differ between reverberant and anechoic environments, and this mismatch between the training data and real data is mitigated by the reverberant augmented training set.
Another interesting derivation is that the RIR model introduced by Valeau \cite{valeau2006use} consistently leads to the highest performance, relative to competing RIR models, for all VADs and in all tested acoustics. One explanation is that the model proposed in \cite{valeau2006use} predicts room acoustics better than the remaining models.
It should be noticed that training with Valeau RIR model leads to rapid convergence of the ROC curves and leading AUC values. These results indicate that detectors trained with Valeau impulse response achieve wide margins of separation between speech and silence. Therefore, these detectors experience high robustness from noises and interferences that might shift the classifier.

\begin{figure}[ht!]
	\centering\includegraphics[width=\columnwidth]{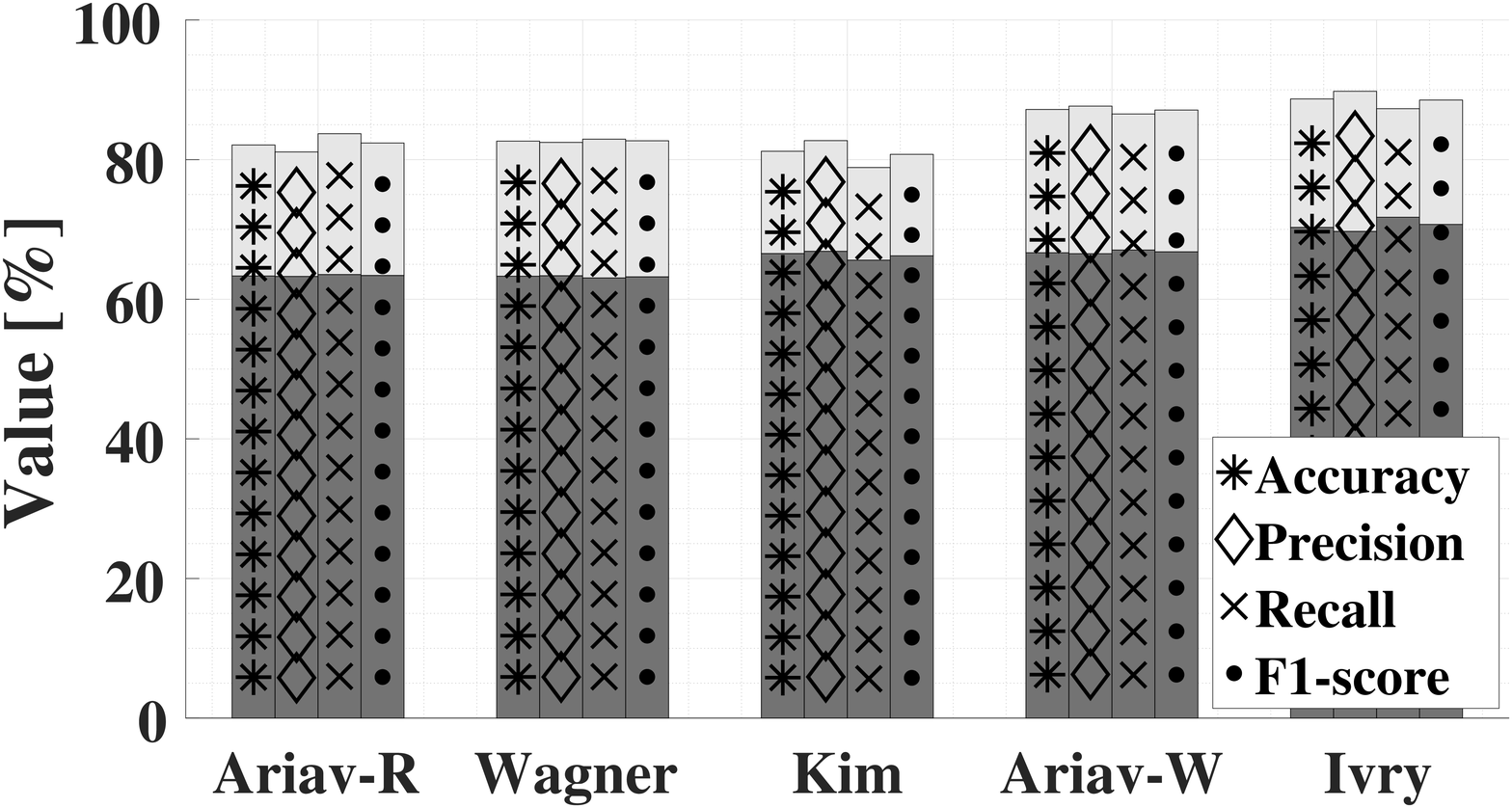}
	\vfill
	\includegraphics[width=\columnwidth]{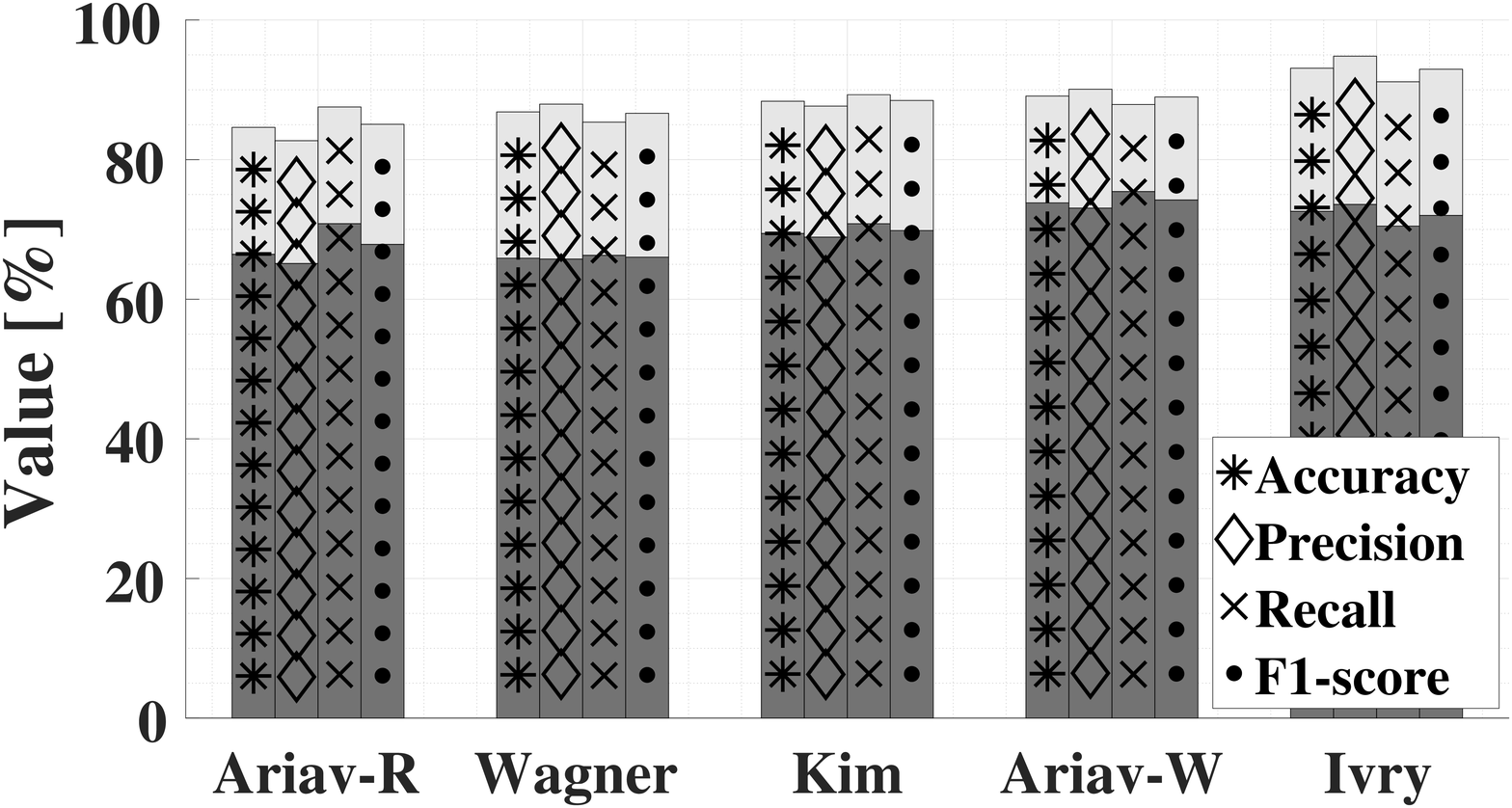}
	\vfill
	\includegraphics[width=\columnwidth]{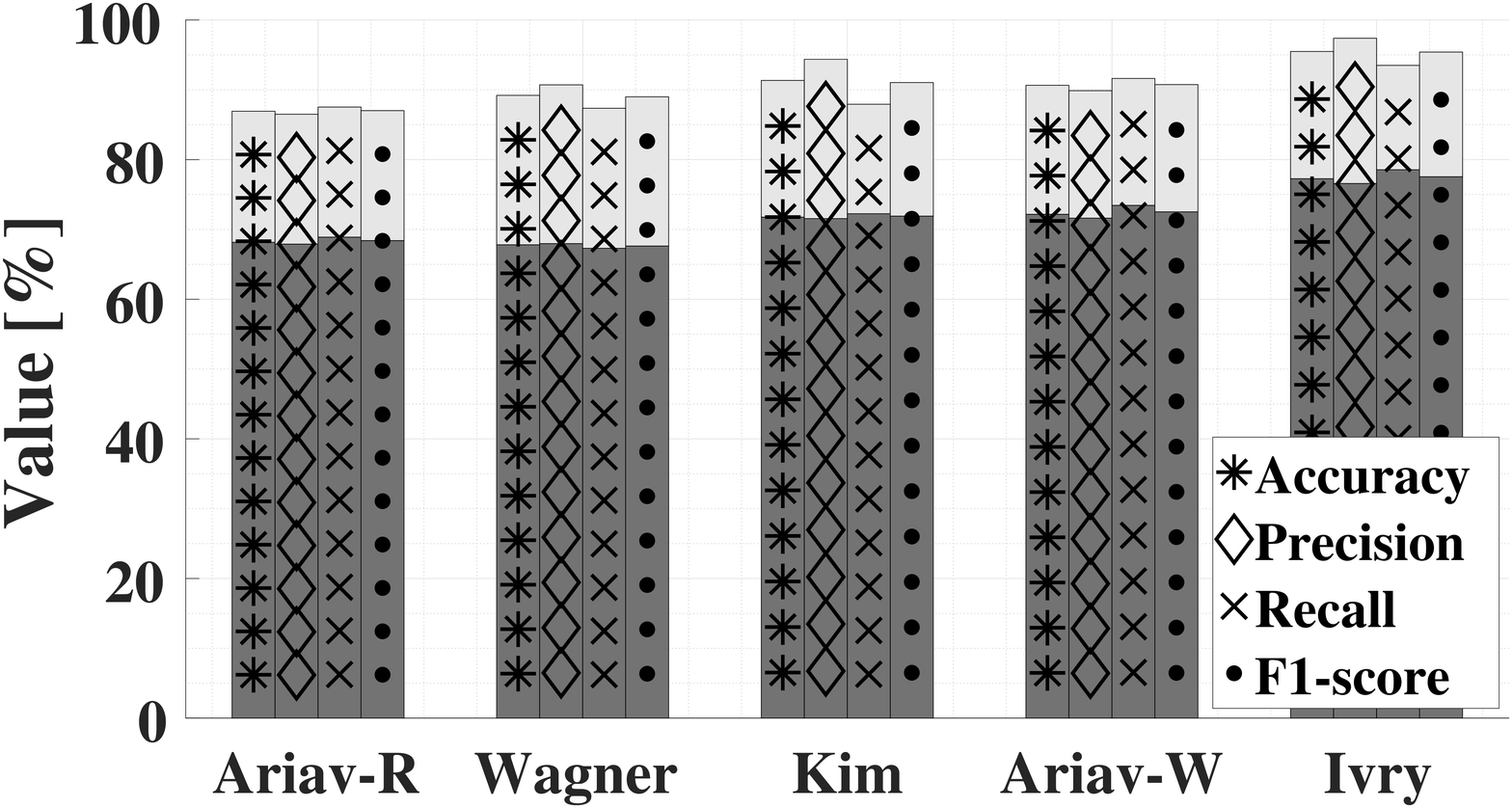}
	\caption{Performance of the five VADs in real reverberant conditions of (top to bottom): classroom, large concert hall, octagon library. Comparison is made between employing anechoic training (dark) and augmented training with Valeau RIR model (light).}
	\label{measures}
\end{figure}

\begin{figure}[ht!]
	\centering\includegraphics[width=\columnwidth]{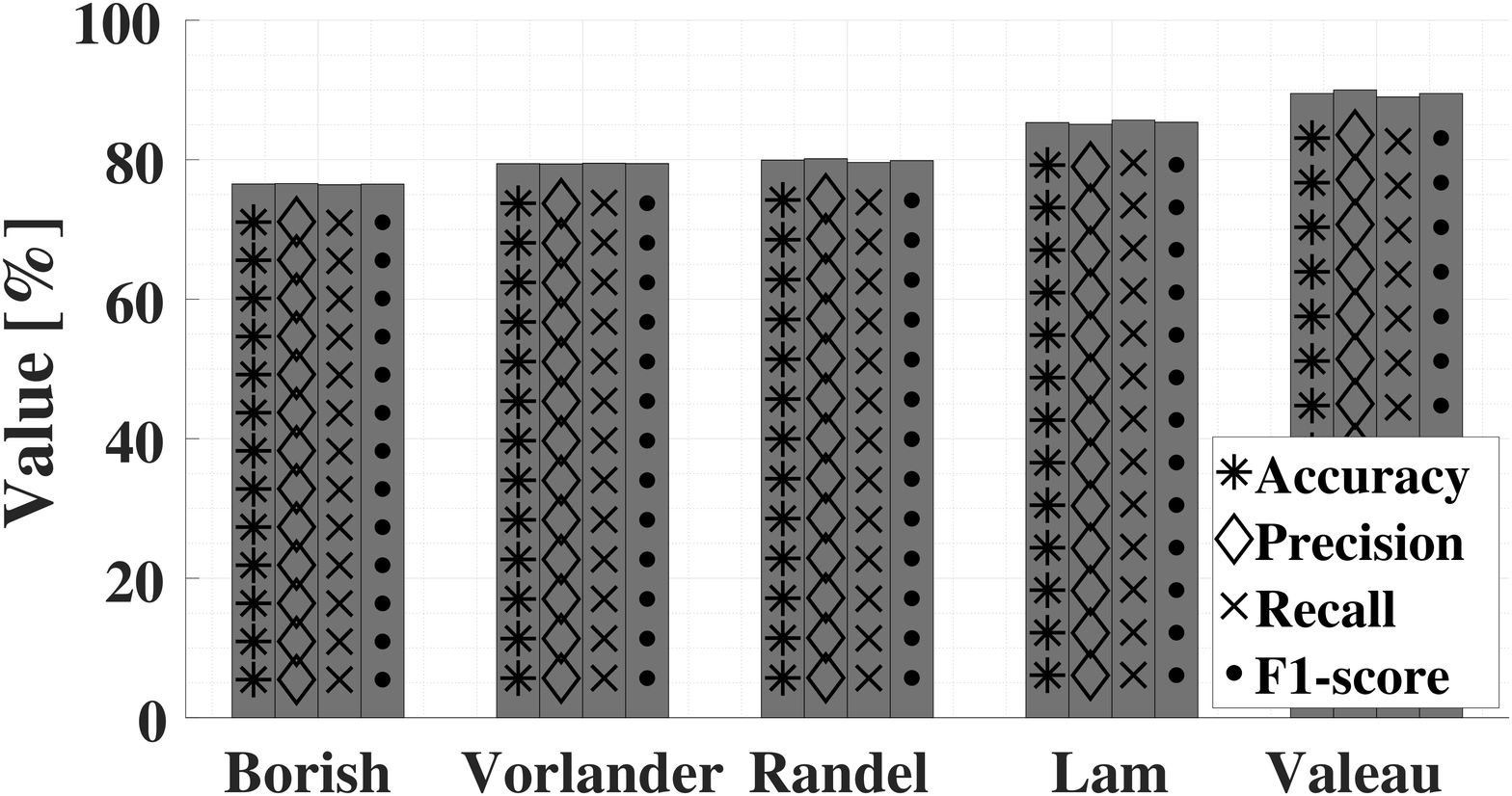}
	\vfill
	\includegraphics[width=\columnwidth]{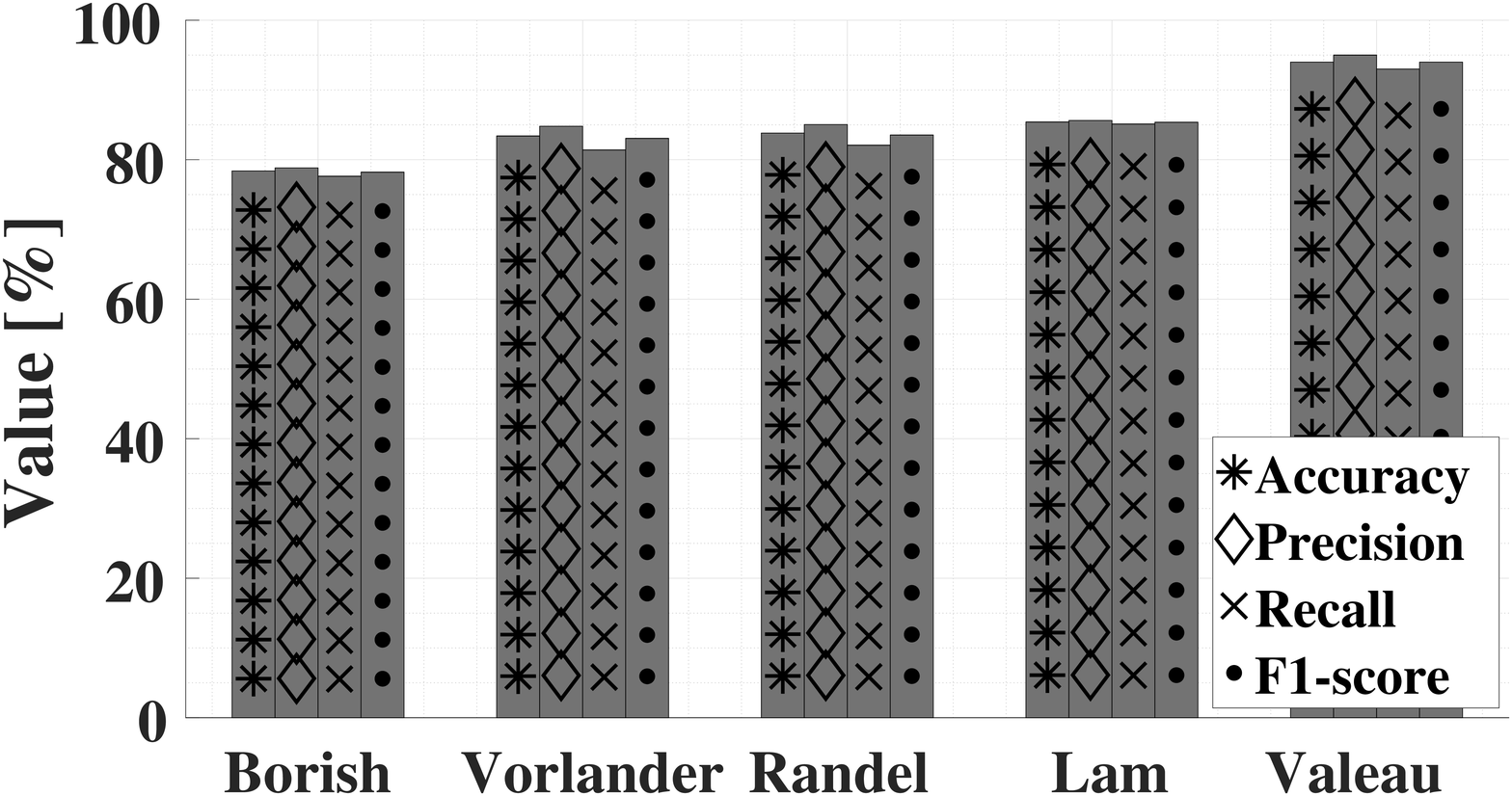}
	\vfill
	\includegraphics[width=\columnwidth]{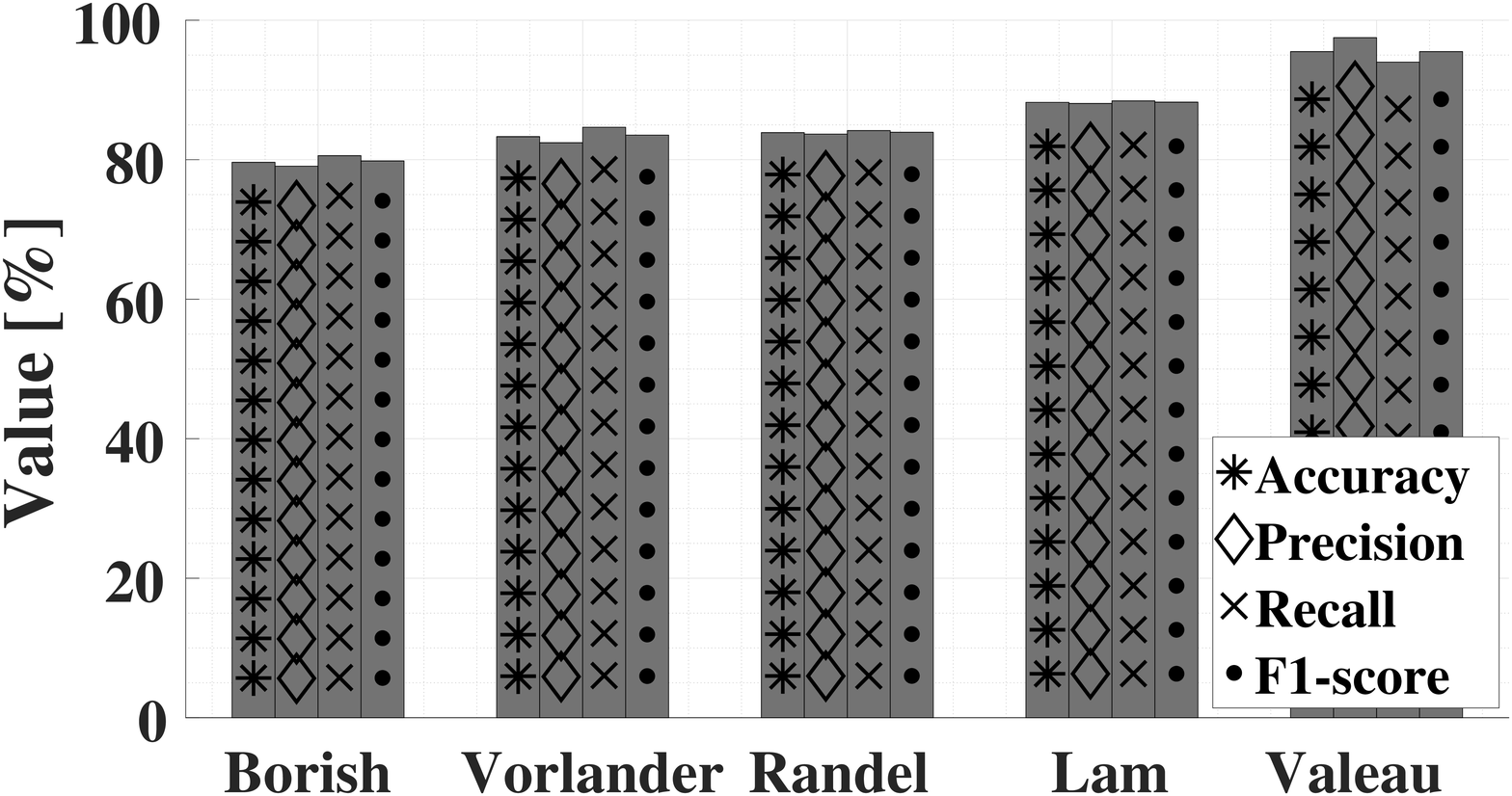}
	\caption{Performance of Ivry VAD \cite{ivry2019voice} in real reverberant conditions of (top to bottom): classroom, large concert hall, octagon library. Comparison is made between the five RIR training models.}
	\label{measuresRIR}
\end{figure}

Further derivations can be made based on Fig \ref{measures}. The reported results reaffirm that augmentation of the training set with respect to Valeau RIR model leads to enhanced VAD performance in reverberant conditions, compared to training that merely considers anechoic data. This enhancement can be quantified by approximately $20\%$ gap across all performance measures of accuracy, precision, recall and F1-score.
This conclusion also implies high generalization ability of all VADs that are trained with \cite{valeau2006use}, since they consistently achieve enhanced performance for all measures and in all three acoustic environments.
Next, let us focus on the interpretation of the accuracy, precision and recall measures. Since the training and test sets are balanced, these values strongly characterize the capabilities of the detector. The accuracy measure confirms that the Valeau model leads to accurate detection in frames of both speech and silence. Also, the enhanced precision measure correspondingly lowers the false-positive value, i.e., non-speech frames have lower probability of being classified as speech. This result highly benefits applications such as speech enhancement, in which interferences may lead to severe degradation in practical performance.
In a similar manner, the increase in recall decreases the false-negative measure. Thus, loss of information that typically lies in speech frames is obviated with higher probability.

Next, we focus on Ivry VAD \cite{ivry2019voice} that achieved leading performance in all previous experiments, as can be viewed in Figs. \ref{ROCs} and \ref{measures}. A further analysis and evaluation of Ivry detector was conducted, and the results are depicted in Fig. \ref{measuresRIR}. This detector obtains a state-of-the-art performance of $95\%$ on average in all reported evaluation measures when trained with Valeau RIR model, which prevails competing VAD methods. Also, this detection system obtains leading accuracy, precision and recall measures across all tested reverberant setups. This outcome points on high generalization ability, robustness for noises and interferences, and prime accuracy in correctly distinguishing speech from silence.

\section{Conclusions} \label{sec:conclusions}
In this study, we have considered five different state-of-the-art deep-learning-based VADs. We have shown that these detectors, when trained with merely anechoic data, experience substantial degradation in performance when tested in reverberant conditions. To mitigate the mismatch between training data and real data, we simulated an augmented training set that contains both an anechoic corpus and its reverberant transformation, where the latter was generated using a fixed room impulse response model. This extension permitted detectors to learn unique patterns and audio-based features that represent reverberant settings. The experiment was performed independently with five different room impulse response models. The training augmentation led to enhanced performance of all VAD systems when tested in three different real-life reverberant spaces of a classroom, a large concert hall and an octagon shaped library. Improvement was obtained in terms of both accuracy, generalization and robustness abilities. Also, evaluation of performance measures yielded an average increase of $20\%$ in accuracy, precision and recall with respect to non-reverberant training corpus. This study has also shown that the response model introduced by Valeau \cite{valeau2006use} consistently leads to the best performance, regardless of the detector and the tested acoustic environment. That and more, the VAD introduced by Ivry \cite{ivry2019voice} has achieved leading performance across all experiments. In future work, additional aspects such as feature engineering and dedicated architecture will be addressed in order to further enhance Ivry detector and adjust it for practical and reverberant acoustic scenarios.

\bibliographystyle{IEEEbib}
\bibliography{refs}

\begin{thebibliography}{10}

\bibitem{kinoshita2016summary}
Keisuke Kinoshita, Marc Delcroix, Sharon Gannot, Emanu{\"e}l~AP Habets,
  Reinhold Haeb-Umbach, Walter Kellermann, Volker Leutnant, Roland Maas,
  Tomohiro Nakatani, Bhiksha Raj, et~al.,
\newblock ``A summary of the {REVERB} challenge: state-of-the-art and remaining
  challenges in reverberant speech processing research,''
\newblock {\em EURASIP Journal on Advances in Signal Processing}, vol. 2016,
  no. 1, pp. 7, 2016.

\bibitem{kuklasinski2016maximum}
Adam Kuklasi{\'n}ski, Simon Doclo, S{\o}ren~Holdt Jensen, and Jesper Jensen,
\newblock ``Maximum likelihood {PSD} estimation for speech enhancement in
  reverberation and noise,''
\newblock {\em IEEE Transactions on Audio, Speech, and Language Processing},
  vol. 24, no. 9, pp. 1599--1612, 2016.

\bibitem{zhao2016dnn}
Yan Zhao, DeLiang Wang, Ivo Merks, and Tao Zhang,
\newblock ``{DNN}-based enhancement of noisy and reverberant speech,''
\newblock in {\em IEEE International Conference on Acoustics, Speech and Signal
  Processing (ICASSP)}. IEEE, 2016, pp. 6525--6529.

\bibitem{han2015learning}
Kun Han, Yuxuan Wang, DeLiang Wang, William~S Woods, Ivo Merks, and Tao Zhang,
\newblock ``Learning spectral mapping for speech dereverberation and
  denoising,''
\newblock {\em IEEE Transactions on Audio, Speech, and Language Processing},
  vol. 23, no. 6, pp. 982--992, 2015.

\bibitem{schwarz2015coherent}
Andreas Schwarz and Walter Kellermann,
\newblock ``Coherent-to-diffuse power ratio estimation for dereverberation,''
\newblock {\em IEEE Transactions on Audio, Speech, and Language Processing},
  vol. 23, no. 6, pp. 1006--1018, 2015.

\bibitem{ko2017study}
Tom Ko, Vijayaditya Peddinti, Daniel Povey, Michael~L Seltzer, and Sanjeev
  Khudanpur,
\newblock ``A study on data augmentation of reverberant speech for robust
  speech recognition,''
\newblock in {\em IEEE International Conference on Acoustics, Speech and Signal
  Processing (ICASSP)}. IEEE, 2017, pp. 5220--5224.

\bibitem{giri2015improving}
Ritwik Giri, Michael~L Seltzer, Jasha Droppo, and Dong Yu,
\newblock ``Improving speech recognition in reverberation using a room-aware
  deep neural network and multi-task learning,''
\newblock in {\em IEEE International Conference on Acoustics, Speech and Signal
  Processing (ICASSP)}. IEEE, 2015, pp. 5014--5018.

\bibitem{ariav2018deep}
Ido Ariav, David Dov, and Israel Cohen,
\newblock ``A deep architecture for audio-visual voice activity detection in
  the presence of transients,''
\newblock {\em Signal Processing}, vol. 142, pp. 69--74, 2018.

\bibitem{wagner2018deep}
Johannes Wagner, Dominik Schiller, Andreas Seiderer, and Elisabeth Andr{\'e},
\newblock ``Deep learning in paralinguistic recognition tasks: Are hand-crafted
  features still relevant?,''
\newblock in {\em Interspeech}, 2018, pp. 147--151.

\bibitem{kim2018voice}
Juntae Kim and Minsoo Hahn,
\newblock ``Voice activity detection using an adaptive context attention
  model,''
\newblock {\em IEEE Signal Processing Letters}, vol. 25, no. 8, pp. 1181--1185,
  2018.

\bibitem{ariav2019end}
Ido Ariav and Israel Cohen,
\newblock ``An end-to-end multimodal voice activity detection using {W}ave{N}et
  encoder and residual networks,''
\newblock {\em IEEE Journal of Selected Topics in Signal Processing}, vol. 13,
  no. 2, pp. 265--274, 2019.

\bibitem{ivry2019voice}
Amir Ivry, Baruch Berdugo, and Israel Cohen,
\newblock ``Voice activity detection for transient noisy environment based on
  diffusion nets,''
\newblock {\em IEEE Journal of Selected Topics in Signal Processing}, vol. 13,
  no. 2, pp. 254--264, 2019.

\bibitem{allen1979image}
Jont~B Allen and David~A Berkley,
\newblock ``Image method for efficiently simulating small-room acoustics,''
\newblock {\em The Journal of the Acoustical Society of America}, vol. 65, no.
  4, pp. 943--950, 1979.

\bibitem{borish1984extension}
Jeffrey Borish,
\newblock ``Extension of the image model to arbitrary polyhedra,''
\newblock {\em The Journal of the Acoustical Society of America}, vol. 75, no.
  6, pp. 1827--1836, 1984.

\bibitem{vorlander1989simulation}
Michael Vorl{\"a}nder,
\newblock ``Simulation of the transient and steady-state sound propagation in
  rooms using a new combined ray-tracing/image-source algorithm,''
\newblock {\em The Journal of the Acoustical Society of America}, vol. 86, no.
  1, pp. 172--178, 1989.

\bibitem{rindel1993modelling}
Jens~H Rindel,
\newblock ``Modelling the angle-dependent pressure reflection factor,''
\newblock {\em Applied Acoustics}, vol. 38, no. 2-4, pp. 223--234, 1993.

\bibitem{lam2005issues}
Yiu~Wai Lam,
\newblock ``Issues for computer modelling of room acoustics in non-concert hall
  settings,''
\newblock {\em Acoustical science and technology}, vol. 26, no. 2, pp.
  145--155, 2005.

\bibitem{valeau2006use}
Vincent Valeau, Judica{\"e}l Picaut, and Murray Hodgson,
\newblock ``On the use of a diffusion equation for room-acoustic prediction,''
\newblock {\em The Journal of the Acoustical Society of America}, vol. 119, no.
  3, pp. 1504--1513, 2006.

\bibitem{garofolo1988getting}
John~S Garofolo, Lori~F Lamel, William~M Fisher, Jonathan~G Fiscus, and David~S
  Pallett,
\newblock ``Getting started with the {DARPA TIMIT CD-ROM}: An acoustic phonetic
  continuous speech database,''
\newblock {\em National Institute of Standards and Technology (NIST),
  Gaithersburgh, MD}, vol. 107, pp. 16, 1988.

\bibitem{stewart2010database}
Rebecca Stewart and Mark Sandler,
\newblock ``Database of omnidirectional and {B}-format room impulse
  responses,''
\newblock in {\em IEEE International Conference on Acoustics, Speech and Signal
  Processing (ICASSP)}. IEEE, 2010, pp. 165--168.

\bibitem{powers2011evaluation}
David~Martin Powers,
\newblock ``Evaluation: from precision, recall and {F}-measure to {ROC},
  informedness, markedness and correlation,''
\newblock {\em Journal of Machine Learning Technologies}, vol. 2, no. 1, pp.
  37--63, 2011.

\end{thebibliography}

\end{document}